\documentclass[aps,twocolumn,fleqn]{revtex4}

\usepackage{graphicx}
\usepackage{amsmath}
\usepackage{amssymb}
\usepackage{amsfonts}
\usepackage{xcolor}
\usepackage{mathbbol}
\usepackage{stmaryrd}
\usepackage{braket}
\usepackage[charter]{mathdesign}

\definecolor{gray}{rgb}{0.5,0.5,0.5}
\definecolor{darkblue}{rgb}{0,0,0.5}
\definecolor{darkred}{rgb}{0.5,0,0}
\definecolor{darkgreen}{rgb}{0,0.5,0}

\DeclareSymbolFontAlphabet{\amsmathbb}{AMSb}

\newcommand{\rmi}{\mathrm{i}}

\def\op#1{\mathbb{#1}}
\def\opH{\op{H}}
\def\opL{\op{L}}
\def\opA{\op{A}}

\def\opJ{\op{J}}
\def\opz{\{ \op{Z}_i \}}
\def\opS{\op{S}}
\def\opX{\op{X}}
\def\opP{\op{P}}
\def\opone{\op{1}}
\def\su#1{{\cal #1}}
\def\suL{\su{L}}

\begin{document}
\title{Symmetry-induced decoherence-free subspaces}

\author{Jonathan Dubois}
\author{Ulf Saalmann}
\author{Jan Michael Rost}
\affiliation{Max Planck Institute for the Physics of Complex Systems, N\"{o}thnitzer Stra{\ss}e 38, 01187 Dresden, Germany}

\begin{abstract}\noindent
Preservation of coherence is a fundamental yet subtle phenomenon in open systems. We uncover its relation to symmetries respected by the system Hamiltonian and its coupling to the environment. We discriminate between local and global classes of decoherence-free subspaces for many-body systems  through the introduction of ``ghost variables''. The latter are orthogonal to the symmetry and the coupling to the environment does not depend on them. Constructing them is facilitated in classical phase space and can be transferred to quantum mechanics through the equivalent role that Poisson and Lie algebras play for symmetries  in classical and quantum mechanics, respectively. Examples are given for an interacting spin system.
\end{abstract}

\maketitle

\paragraph*{Introduction:}
A physical system  interacting with an environment~\cite{Gardiner1983, Risken1984, Gotze1992, Breuer2002} relaxes to an equilibrium or a non-equilibrium steady state on time scales much longer than the relaxation time~\cite{Qian2006}. In the steady state  the  system no longer evolves in time due to its coupling to the environment and \emph{all} associated physical observables reach a constant value. This characteristic is inherent to classical and quantum open systems.

Quantum mechanically, relaxation is accompanied by decoherence which remains a major obstacle in putting physical devices to work for quantum computation~\cite{Grover1997, Ladd2010, nich10}. 
Therefore,  decoherence-free subspaces~\cite{Lidar1998, Albert2016, Tarnowski2021} (DFS), which are protected against decoherence effects~\cite{Habib1998}, play a crucial role in realizing quantum computing~\cite{Grover1997, Mohseni2003, Friesen2017}.
Understanding and formulating the general  conditions underlying DFS is the prerequisite to tame decoherence and to establish open systems  which exhibit non-stationary long-time dynamics (NLD), that is, remain out of equilibrium despite their interaction with an environment~\cite{Buca2019, Tindall2020, Tarnowski2021}. 

In closed systems, it is well known that symmetries, i.e., constants of motion, are crucial to characterize their dynamics. Here, we will show that this is also the case for open systems: Symmetries can be used to formulate simple conditions for DFS in terms of the dependence of the system and the coupling to the environment on these symmetries. Moreover, these conditions can be identified and formulated in an intuitive way via classical dynamics since exact symmetries hold quantum mechanically as well as classically through their equivalent formulation in terms of commutators and Poisson brackets, respectively.
The classical phase-space perspective will also allow us to introduce naturally new classes of DFS where the coupling to the environment depends only on so called ``ghost variables'', orthogonal to the symmetries. 
Finally, we will present the quantized conditions for the DFS in terms of eigenstates of the symmetry operator.

Our starting point is the Lindblad master equation~\cite{Breuer2002} $\dot{\rho} {=} - \rmi \mathcal{L}(\opH,\opL) \rho$ for the systems's density matrix $\rho$ with Hamiltonian $\opH$ coupled to the the environment in Markov approximation through the Lindblad operators $\opL$. 
In the framework of the master equation DFS exist if and only if the Lindbladian $\mathcal{L}$ has nonzero real eigenvalues~\cite{Lidar1998, Baumgartner2008, Albert2014, Buca2019}. 
Here, we will work with the adjoint Lindbladian $\mathcal{L}^{\dagger}(\opH,\opL)$ which has identical spectral properties, since we are interested in operators, most prominently a symmetry $\opJ$. As a constant of motion for the open system its dynamics is governed in the Heisenberg picture by $\mathcal{L}^{\dagger}$  through
\begin{subequations}\label{eq:adj}
\begin{align}\label{eq:sym}
    0=&\frac{{\rm d} \opJ}{{\rm d} t} = \rmi \mathcal{L}^{\dagger}\opJ,
    \tag{\ref{eq:adj}}
\intertext{with $\mathcal{L}^{\dagger}=\mathcal{L}^{\dagger}_\opH+\mathcal{L}^{\dagger}_\opL$, where}
\label{eq:adj-parts}
    \mathcal{L}^{\dagger}_\opH\opJ = &\dfrac{1}{\hbar} [\opH,\opJ],\\  \mathcal{L}^{\dagger}_\opL\opJ = & \dfrac{\rmi}{\hbar}\sum_\alpha\left(\opL^\dagger_\alpha[\opJ,\opL_\alpha]+[\opL^\dagger_\alpha,\opJ]\opL_\alpha\right)\,. \label{eq:adj-parts-L}
\end{align}
\end{subequations}
Since symmetries are respected equivalently by a quantum system and its classical counterpart (here, via Poisson brackets), we can give an intuitive account of these conditions in classical phase space~\cite{Strunz1998, Dubois2021}. For a constant of motion $J$ with $\lbrace H , J \rbrace \,{=}\, \lbrace L_{\alpha} , J \rbrace \,{=}\, 0$,  we find that semiclassically DFS exist if there is at least one layer $J {=} J_0$ on which 
\begin{equation}
    \label{eq:NLD_classical}
    \omega (J_0) \equiv \left.\partial H/\partial J\right|_{J_0} \neq 0, 
    \quad  
    \ell_{\alpha} (J_0) \equiv \left.\partial L_{\alpha}/\partial J\right|_{J_0} = 0, 
\end{equation}
where $\omega (J)$ is the Hamiltonian frequency and $\ell_{\alpha} (J)$ can be interpreted as the decay rate induced by the environment. Note, that we denote quantum operators with $\opX$  in contrast to (classical) functions $X$. While the condition on $H$ is necessary for (oscillatory) non-stationary dynamics in the first place, the condition on $L_{\alpha}$ makes sure that this dynamics is preserved for long times and therefore establishes NLD. Interestingly, since $L_{\alpha}{=}L_{\alpha}(J,\mathbf{Z})$ is in principle a function of all phase-space variables (boldface letters denote a set of variables), the condition $\ell_{\alpha} (J){=}0$ can be achieved in two, qualitatively different ways:
\begin{itemize}
    \item[(i)]
    local realization for a specific $J_0$: the $L_{\alpha}$ depend explicitly on $J$ but their derivative  vanishes for $J_0$; 
    \item[(ii)]
     global realization for all $J$: the $L_{\alpha}$ do not depend on $J$ but can depend on all the variables $\mathbf{Z}$, the \emph{ghost variables}, being orthogonal to $J$.
\end{itemize}
These two ways exist also quantum mechanically, where the conditions on the symmetry \eqref{eq:NLD_classical} are quantized in the form that there exist two eigenspaces $\{|n\rangle\}$ and $\{|m\rangle\}$ of $\opJ$ with eigenvalues $J_n$ and $J_m$, respectively, such that
\begin{equation}\label{eq:NLD_qm}
     \Delta\opH_{nm} \ne 0\quad \mbox{and}\quad \Delta\opL_{\alpha nm}=0,
 \end{equation}
with the explicit forms of $\Delta\opH_{\alpha nm}$ and $\Delta\opL_{\alpha nm}$ given below in Eq.\,\eqref{eq:Lindbladian_block}.

\begin{figure}
    \centering
    \includegraphics[width=\columnwidth]{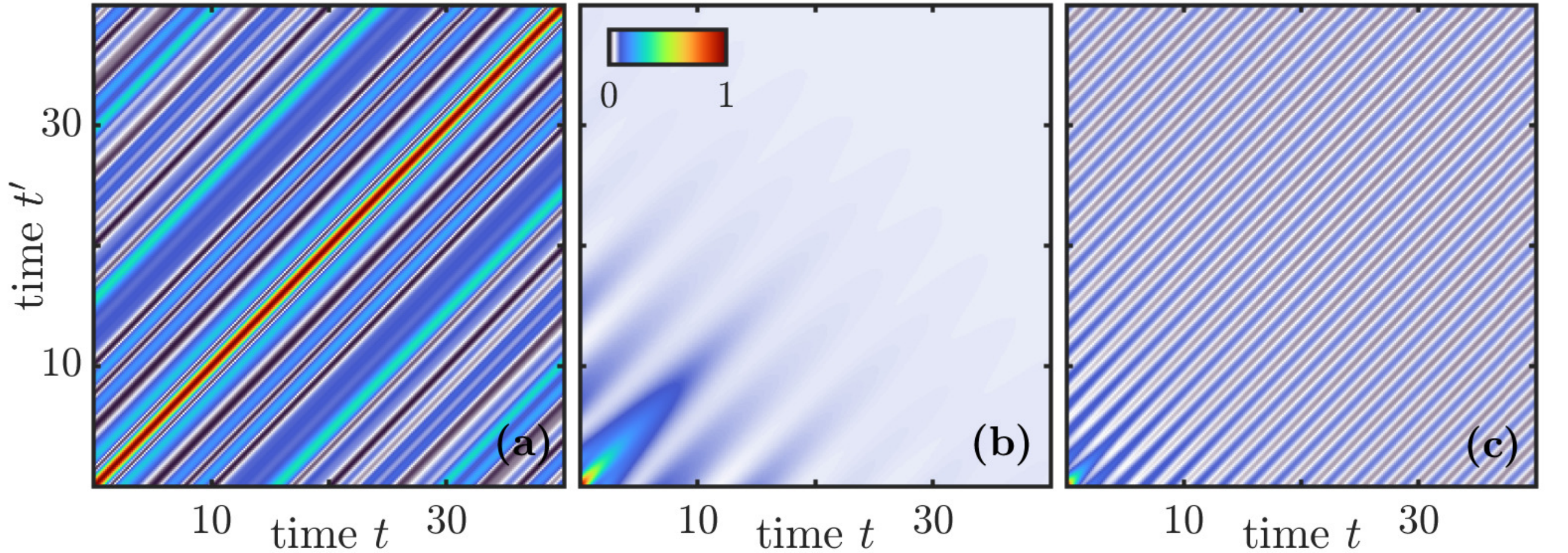}
    \caption{Autocorrelation function ${\rm tr} \rho (t) \rho (t^{\prime})$ for Hamiltonian~\eqref{eq:Heisenberg_model} with $N{=}6$. Parameters $\omega_i$, $\delta_{ij}$ and $\Delta_{ij}$ are given in the supplement~\cite{SuppMat}. (a) No dissipation, $\mathcal{L}_\opL = 0$. (b) Dissipation with $N$ Lindblad operators $\opL_\alpha = \gamma \opS_\alpha^z$. (c) Dissipation with $(N{-}1)N/2$ ghost operators such that $\opL_{\alpha_{jk}} = \gamma (\opS_j^z - \opS_k^z)$. In both cases $\gamma {=} 1/3$.}
    \label{fig:autocorrelation}
\end{figure}

Since systems which we classify now as local realizations (i) have been discussed in~\cite{Buca2019, Tindall2020}, we  focus on global realizations (ii) in the following. Figure~\ref{fig:autocorrelation} shows such a global realization with a Heisenberg spin model in comparison to purely Hamiltonian dynamics (Fig.\,\ref{fig:autocorrelation}a) and to the case with an arbitrary dissipative part (Fig.\,\ref{fig:autocorrelation}b). 
Indeed, for an environment depending on  ghost operators only (Fig.\,\ref{fig:autocorrelation}c), coherent oscillations are preserved despite dissipation.
Before we construct the ghost operators explicitly via the corresponding classical ghost variables, we derive the conditions \eqref{eq:NLD_classical}.

Consider phase-space variables $(J,\theta,\mathbf{Z})$ with $J$  conserved,  $\lbrace J,H\rbrace \,{=}\, \lbrace J,L_{\alpha}\rbrace \,{=}\, 0$. The angle $\theta$ is canonically conjugate to the action-like $J$ such that $\lbrace \theta , J \rbrace \,{=}\, 1$. Then, the phase space is foliated by manifolds on which $J$ is conserved. The  $\mathbf{Z}$ are a set of canonically (or noncanonically) conjugate variables such that $\lbrace \mathbf{Z} , J \rbrace \,{=}\, \lbrace \mathbf{Z} , \theta \rbrace \,{=}\, \boldsymbol{0}$. The relevance of the \emph{ghost variables} $\mathbf{Z}$  for DFS will  become clear below. 

From the action-angle variables  we can construct ``beating variables''
\begin{equation}\label{eq:beating}
    A = f(J) \exp (-\rmi \theta)\,\,\,\,\text{and}\,\,\,\,A^* = f(J) \exp (\rmi \theta ) \, ,
\end{equation}
with $f (J) = A^{*} A$~\cite{Campoamor2012, SuppMat}. The beating variables describe dynamics  along a path on which $J$ is conserved. 
In the semiclassical limit of the adjoint Lindbladian~\cite{Strunz1998, Dubois2021, SuppMat} the dynamics of $A$ on the manifold of constant $J$ is given by $\dot{A} = \rmi \mathcal{L}^{*} A$ with
\begin{equation}
    \label{eq:A_dynamics}
    \mathcal{L}^{*} A = {-} \bigg(\partial_J H + \sum_{\alpha} \Big( 2 {\rm Im} \, ( L_{\alpha} \partial_J L_{\alpha}^{*} )  - \rmi \hbar|\partial_J L_{\alpha}|^2 \Big) \bigg)\, A,
\end{equation}
which follows from the Poisson brackets $\lbrace A , H \rbrace \,{=}\, {-}\rmi(\partial_J H)\, A$ and $\lbrace A , L_{\alpha} \rbrace \,{=}\, {-}\rmi(\partial_J L_{\alpha})\, A$, where $\partial_J H$ is real and $\partial_J L_{\alpha}$ can be complex (since $L_{\alpha}$ can be complex). 
Hence, $A$ evolves in time with an oscillatory and a decaying part, regardless of the dynamics of the ghost variables $\mathbf{Z}$.  The oscillatory part comes from a combination of the Hamiltonian and the dissipative components, while the decay part is a consequence of  dissipation only. Note, that the decaying part is of order $\hbar$ and is a consequence of diffusion or quantum noise.

If $\partial_J L_{\alpha}{=}0$ the decaying part in \eqref{eq:A_dynamics} vanishes and the beating variable $A$ exhibits NLD, provided that $\partial_J H\ne 0$ which leads to the conditions \eqref{eq:NLD_classical}  specifying the existence of DFS.
If the Lindblad function depends explicitly on the conserved quantity $J$ and there exists a manifold $J {=} J_0$ on which $\ell_{\alpha} (J_0,\mathbf{Z}) {=} 0$, the environment couples to the degree of freedom associated with $J$, but not on the subspace $J{=}J_0$. Although $J$ is conserved, NLD is restricted to the manifold $J {=} J_0$, and therefore we call this realization (i) of DFS \emph{local}.
One also gets $\ell_{\alpha}(J,\mathbf{Z}) {=} 0$, if $L_{\alpha}$ does not depend explicitly on  $J$ at all. In this case, the environment does not couple to the degree of freedom associated with $J$. These conditions apply to all manifolds labeled by $J$, and therefore this constitutes a \emph{global} realization (ii) of DFS.

In both cases, the environment does not affect the oscillatory dynamics of the beating variable $A$  characterized by the (non-zero)  real eigenvalues $\omega (J,\mathbf{Z})$ of the Lindbladian. If the Hamiltonian frequency $\omega$ depends explicitly on the ghost variables $\mathbf{Z}$, this oscillatory dynamics can be very complicated. If  $\omega {=} \omega(J)$ only, one can directly solve $\dot{A} = \rmi \mathcal{L}^{*} A$ to obtain  $A(t) = \exp (-\rmi \omega(J) t ) A(0)$. In case (i) the environment extinguishes for long times all oscillatory motion linked to $J$ but one with  frequency $\omega(J_0)$ on the manifold $J_0$.

From the semiclassical perspective as developed above we can draw further conclusions.
The conditions~\eqref{eq:NLD_classical} do not require integrability (or  near-integrability) of the system. Therefore, also classically chaotic systems can have DFS, which we demonstrate here explicitly with the Heisenberg spin model. Furthermore, if DFS exist due to a symmetry, small perturbations of the system will not destroy them. This follows from the KAM theorem~\cite{delaLlave2005}, which ensures the existence of modified action-angle variables for open systems subjected to perturbations.
Therefore, we construct now explicitly a coupling to the environment according to (ii) for a chaotic  Hamiltonian.

\paragraph*{Application to the Heisenberg XXZ spin model:}

\begin{figure}
    \centering
    \includegraphics[width=0.333\textwidth]{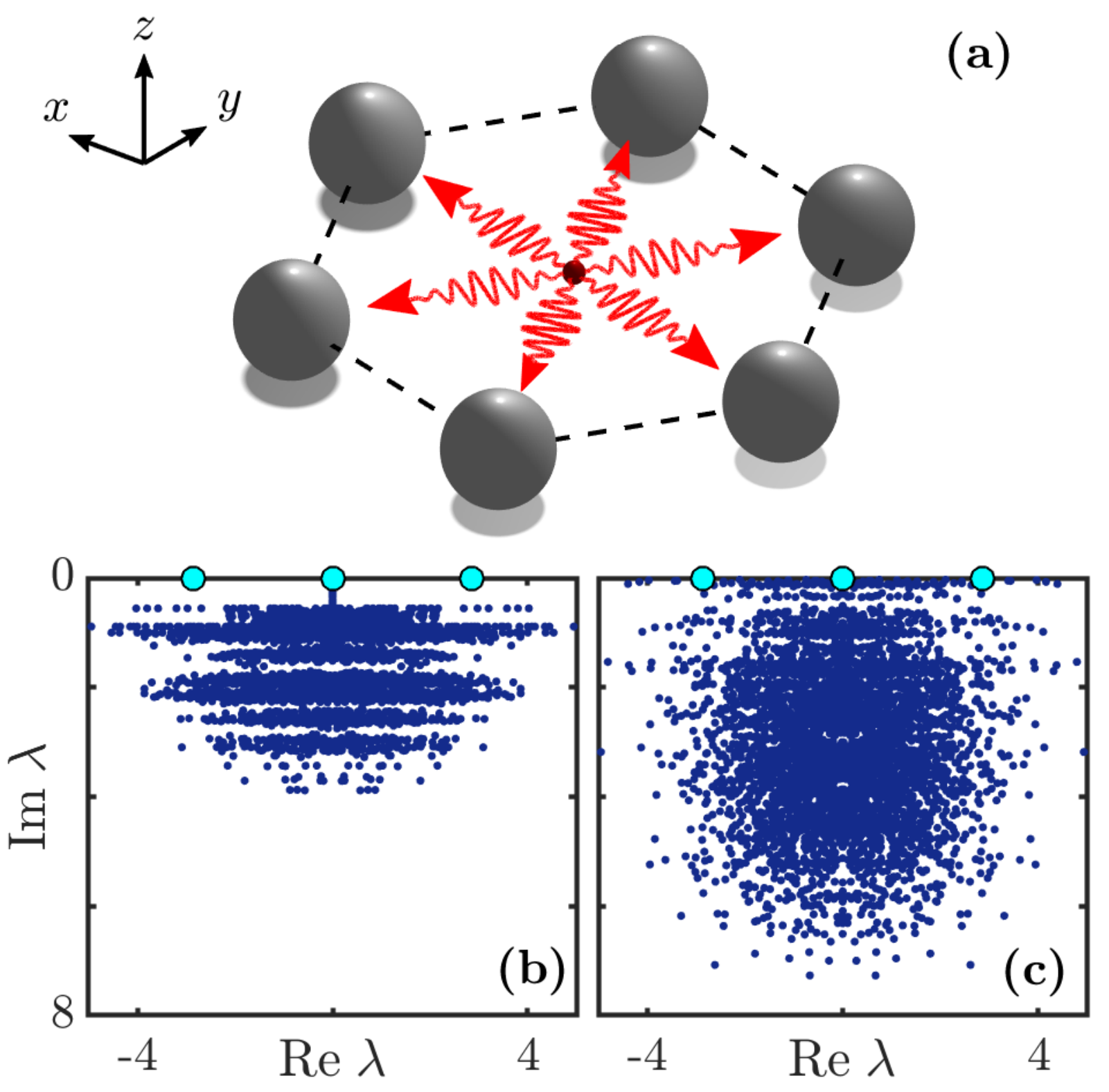}
    \caption{(a) Schematic of the spin system for $N {=} 6$ governed by Hamiltonian~\eqref{eq:Heisenberg_model} as used before in Fig.\,\ref{fig:autocorrelation}. The red lines illustrate the dissipations induced by the ghost operators. (b) and (c) Spectrum of the adjoint of the Lindbladian $\mathcal{L}^{\dagger}$. The cyan circles are the purely real eigenvalues. (b) For $(N{-}1)N/2$ Lindblad operators as a linear combination of the ghost operators~\eqref{eq:ghosts_Sz} such that $\opL_{\alpha_{ij}} = \gamma ( \opS_i^z - \opS_j^z )$. (c) For $N{-}1$ Lindblad operators as a function of the ghosts~\eqref{eq:ghosts_Sbeating} with $\opL_{\alpha} = \gamma \widetilde{\opS}_\alpha^+$. In both cases $\gamma{=}1/3$.}
    \label{fig:Heisenberg}
\end{figure}

A realization of (ii) requires the Lindblad operators $\opz$ to be independent of the symmetry $\opJ$ of the system.
Therefore, only many-body systems render this case nontrivial.
We consider the Heisenberg XXZ spin model~\cite{Prosen2011, Ganahl2012, Liu2014, Sa2020}
\begin{equation}
\label{eq:Heisenberg_model}
    \opH = \sum_{i} \omega_i \opS^z_i + \sum_{ij} \left[ \sigma_{ij} \left( \opS^{+}_i \opS^{-}_j {+} \opS^{-}_i \opS^{+}_j \right) + \Delta_{ij} \opS^z_i \opS^z_j \right] ,
\end{equation}
where the $\mathbf{\opS}_i {=} (\opS^x_i,\opS^y_i,\opS^z_i)$ are spin $1/2$ operators on site $i$ from Pauli matrices, and $\opS^{\pm}_i = \opS^x_i \pm \rmi\,\opS^y_i$ are ladder operators, with $i = 1,...,N$ and $N$ the number of spins,
evoking spin flips between site $i$ and $j$ in the combinations $\opS^{\pm}_i \opS^{\mp}_j$.
The total spin $\opJ = \opS^z = \sum_{k{=}1}^{N} \opS^z_k / N$ is conserved in time as $\mathcal{L}^{\dagger}\opJ {=} 0$ and this leads to pertinent oscillations associated with $\opJ$, see Fig.~\ref{fig:autocorrelation}c.
The parameters used in Eq.\,\eqref{eq:Heisenberg_model} are given in the supplemental material~\cite{SuppMat}.

We construct now an environment depending on ghost operators $\opz$ only which does not affect these oscillations. First, we derive classical ghost variables  with canonical transformations and transfer the ghost variables to ghost operators in the quantum domain.
To this end, we introduce polar coordinates   $\mathbf{S}_i = [ S_i^2-(S^z_i)^2 ]^{1/2} (\mathbf{e}_x \cos \theta_i + \mathbf{e}_y \sin \theta_i) + \mathbf{e}_z S^z_i$ for the spin on site $i$, where the azimuthal angles  $\theta_i$ are canonically conjugate to the  $S^z_i$  and $S_i{}^2 {=} |\mathbf{S}_i|^2$. 
Obviously, Hamiltonian~\eqref{eq:Heisenberg_model} is invariant under rotation with arbitrary azimuthal angle  and therefore $J = S^z / N= \sum_i S^z_i / N$ is conserved. 
Akin to relative center-of-mass motion for massive particles we take advantage of this fact by introducing
relative angles $\Delta \theta_k = \theta_{k{+}1}-\theta_{k}$ for $k = 1,...,N-1$, and the total angle $\theta = \sum_{i} \theta_i$. They are related to the Jacobi coordinates for celestial many-body systems~\cite{Cabral2002}. 

The canonical transformation from phase-space variables $(\theta_i , S^z_i)$ to phase-space variables $(J,\theta,\mathbf{Z} )$, with ghost variables $\mathbf{Z} = ( \Delta \theta_k , \widetilde{S}_k )$ for $k = 1 , ... , N-1$, and its inverse are given in the supplemental material~\cite{SuppMat}. 
The transformation is obtained by using the canonical rules~\cite{Goldstein1980} with a suitable  generating function associated with the center of mass and the relative angles.
Besides $\Delta \theta_k$, we obtain as ghost variables~\cite{SuppMat}
\begin{equation}
    \label{eq:ghost_varibales}
    \widetilde{S}_k^z = \dfrac{k}{N} \sum_{i{=}1}^{N} S^z_i - \sum_{i{=}1}^{k} S^z_i ,
\end{equation}
which are linear functions of the spin variables.
Note that using canonical transformations ensures that the new set of variables spans the entire phase space.
If the interaction with the environment is any combination of these variables or operators (i.e., if it does not depend explicitly on $\theta$ and $J$) DFS exist and the open system exhibits NLD. The ghost variables or operators do not affect the oscillatory dynamics associated with the conserved quantity $J$. Therefore, coherent oscillations persist on long-timescales and the spectrum of the Lindbladian contains nonzero real eigenvalues, cf.\ Fig.~\ref{fig:Heisenberg}. 

The beating variables associated with the conserved quantity, as given in Eq.\,\eqref{eq:beating},
are obtained in terms of spin variables  by performing a non-canonical change of coordinates~\cite{SuppMat} leading to $A \propto \prod_{i{=}1}^{N} S^{-}_i$ and $A^{*} \propto \prod_{i{=}1}^{N} S^{+}_i$. The beating variables oscillate at a frequency $\Omega = \sum_{i{=}1}^{N} \omega_i$, corresponding to the absolute value of the nonzero real eigenvalues and are not affected  by dissipation. 
The \emph{ghost operators} as the quantum analogs of the \emph{ghost variables} follow simply from replacing the Cartesian spin variables in the  ghost variable expressions by the corresponding Cartesian spin operators. Therefore, the ghost operators are functions of the spin operators (Pauli matrices) and read 
\begin{subequations}\label{eq:ghost_operators}
\begin{eqnarray}
    && \widetilde{\opS}_k^z = \dfrac{k}{N} \sum_{i{=}1}^{N} \opS^z_i - \sum_{i{=}1}^{k} \opS^z_i , \label{eq:ghosts_Sz} \\
    && \widetilde{\opS}_k^{\pm} = \mathbb{D}_k \; \opS^{\pm}_{k+1} \, \opS^{\mp}_{k} , \label{eq:ghosts_Sbeating}
\end{eqnarray}
\end{subequations}
with $k = 1,...,N{-}1$ and   ia diagonal matrix $\mathbb{D}_k$~\cite{SuppMat}. 
By construction, all combinations of ghost operators are also ghost operators.
Any Lindbladian  which is an analytic function of these ghost operators leads to DFS.

In Fig.\,\ref{fig:Heisenberg}b, we have used $(N{-}1)N/2$ Lindblad operators as a linear combination of the ghost operators~\eqref{eq:ghosts_Sz} such that $\opL_{\alpha_{jk}} = \gamma ( \opS_j^z - \opS_k^z )$ with $\gamma{=}1/3$ and  
$\alpha_{ij}=(j{-}1)(j{-}2)/2{+}i$ for $j=2,...,N$ and $i=1,...,j{-}1$. 
For Fig.~\ref{fig:Heisenberg}c, we have used $N{-}1$ Lindblad operators of the ghost operators~\eqref{eq:ghosts_Sbeating} such that $\opL_{\alpha} = \gamma \widetilde{\opS}_\alpha^+$ with $\gamma{=}1/3$ and $\alpha = 1,...,N-1$. In both cases, the coherent oscillations associated with the symmetry $\opJ$ are not affected, neither are the beating operators  $\opA = \prod_{i{=}1}^{N} \opS^{-}_i$ and $\opA^{\dagger} = \prod_{i{=}1}^{N} \opS^{+}_i$. Note also, that the (real) eigenvalues (cyan circles) are the same for the cases of  Fig.~\ref{fig:Heisenberg}b and~\ref{fig:Heisenberg}c, since they come from the Hamiltonian part of the Lindbladian $\mathcal{L}_\opH^{\dagger}$ which we have not changed.
We find $\mathcal{L}^{\dagger}_\opH \opA {=} \Omega \opA$ and $\mathcal{L}^{\dagger}_\opL \opA {=} 0$ corresponding to the cyan circles for positive real eigenvalues (the negative one corresponds to the one associated with $\opA^{\dagger}$) in Figs.\,\ref{fig:Heisenberg}b and~\ref{fig:Heisenberg}c.

\paragraph*{Theory of symmetry-induced DFS in quantum mechanics:}
Finally, we are in a position to briefly sketch the derivation of the quantized condition \eqref{eq:NLD_qm} for DFS. We have a set of commuting operators,  $\opz$, its conjugate $\opJ$ and  and the ghost operators. Together, they span the Hilbert space for the open system dynamics. 

The Hamilton operator $\opH$ (and similarly all Lindblad operators $\opL_{\alpha}$) with the symmetry $\opJ$ can be written as a direct sum of irreducible (square) matrices
\begin{equation}
    \label{eq:Hn_Ln}
    \opH = \bigoplus_{n} \opH_n , \qquad \opL_{\alpha} = \bigoplus_{n} \opL_{\alpha n} , 
\end{equation}
where $\opX_n$ is associated with the eigenspace of $J_n$, an eigenvalue of $\opJ$.
They are defined as $\opX_n \,{=}\, {\rm tr}_{n} \opP_n \opX \opP_n$ where $\opP_n \,{=}\, \prod_{j(J_{\!j}\ne J_{n})} ( \opJ{-}J_{\!j}\opone ) / ( J_{n}{-}J_{\!j} )$ are projectors~\cite{Lowdin1962} rendering all subspaces to null operators but the one associated with the eigenvalue $J_n$ of $\opJ$. ${\rm tr}_{n}$, finally, takes the partial trace over all complementary subspaces, reducing the dimension of $\opX_n$ to  the irreducible one of $\opJ_n$. 
Corresponding to the multiplicity $\eta_{n}$ of the eigenvalue $J_{n}$, the $\opH_{n}$ and $\opL_{\alpha n}$ are  $\eta_{n}{\times}\eta_{n}$ matrices.
Due to the block-diagonal form~\eqref{eq:Hn_Ln}, the superoperator $\mathcal{L}^{\dagger}$ can be written as blocks associated with a pair of eigenvalues $(J_n,J_m)$ such that $\mathcal{L}^{\dagger} {=} \bigoplus_{n,m} \mathcal{L}_{nm}^{\dagger}$. Each block of the superoperator
\begin{subequations}\label{eq:Lindbladian_block}
\begin{equation}\label{eq:adj-block}
    \suL^{\dagger}_{nm} =\frac{1}{\hbar}\Delta\opH_{nm} +\frac{\rmi}{\hbar}\sum_{\alpha}\big(\opL^{\dagger}_{\alpha n}\Delta\opL_{\alpha nm} +\Delta\opL^{\dagger}_{\alpha nm}\opL_{\alpha m}\big) 
    \tag{\ref{eq:Lindbladian_block}}
\end{equation}
is of size $(\eta_{n}{\times}\eta_{m}){\times}(\eta_{n}{\times}\eta_{m})$
with \cite{Eins}
\begin{align}
    \label{eq:DHn_def}
    \Delta \opH_{nm} &\equiv \opH_n \otimes \mathbb{1}_m -  \mathbb{1}_n \otimes \opH_m^{\top} , 
    \\ 
    \label{eq:DLn_def}
    \Delta \opL_{\alpha nm} &\equiv \opL_{\alpha n} \otimes \mathbb{1}_m -  \mathbb{1}_n \otimes \opL_{\alpha m}^{\top} .
\end{align}
\end{subequations}
Coherent oscillations of a beating operator $\opA_{nm}$, which is an eigenoperator of $\suL^{\dagger}_{nm}$ with nonzero real eigenvalues, occur for those $\Delta \opH_{nm}$ and $\Delta \opL_{\alpha nm}$ that fulfill condition~\eqref{eq:NLD_qm}.
The eigenoperators $\opA_{nm}$ are matrices with an $\eta_{n}{\times}\eta_{m}$ non-vanishing part such that $\opA_{nm} \,{=}\,\opP_{n}\opA_{nm}\opP_{m}$ and can have multiple beating frequencies.
Furthermore, the $\opA_{nm}$ are also eigenoperators of the Lindbladian since under condition~\eqref{eq:NLD_qm}, with~\eqref{eq:DLn_def}, $[\opL_{\alpha n} , \opL_{\alpha n}^{\dagger}]  \,{=}\, [\opL_{\alpha m} , \opL_{\alpha m}^{\dagger}]  \,{=}\, 0$. The fact that the Lindbladian and its adjoint share the same eigenoperators associated to real eigenvalues show that dissipation plays a passive role in this case.

This completes our classical, semiclassical and quantum treatment of symmetry-induced DFS. It clearly reveals that the notion of non-stationary (oscillating) behavior and its equivalence to real eigenvalues of the Lindbladian applies equally to classical, semiclassical and quantum systems which establishes DFS for classical and semiclassical systems. For the latter, while coherent oscillations in classical systems can still occur in the presence of dissipation (through the second term in~\eqref{eq:A_dynamics}, see~\cite{Keeling2010, Bhaseen2012, Munoz2019} for examples), diffusion can inhibit long-time oscillatory non-stationary motion (through the third term in~\eqref{eq:A_dynamics} being of order $\hbar$). The eigenoperator  of the adjoint Lindbladian corresponding to the real nonzero eigenvalues can be constructed classically, see Eq.\,\eqref{eq:beating}, or quantum mechanically ($\opA_{nm}$) from a symmetry. Playing the role of a beating operator, $A$ or $\opA_{mn}$ connect different elements of the constant of motion $J$ or $\opJ$ in form of its layers in phase space or its irreducible representations in Hilbert space, respectively. Elements $\opJ_n$ of the symmetry itself can be expressed in terms of the beating operator as $\opJ_n \,{=}\, \opA_{mn}^{\dagger} \opA_{mn}$. 
This irreducible representation also plays a crucial role in the context of breaking the ergodicity thermalization hypothesis in Hamiltonian systems~\cite{Sala2020}, which is related to NLD in the absence of dissipation.
For DFS, the Lindblad operators must be degenerate with respect to at least two different irreducible subspaces $n\ne m$ of the symmetry, while the hamiltonian must not be degenerate, see~\eqref{eq:NLD_qm}. This gives a clear picture how DFS emerge. 

Guided by this insight, we have identified a new class of couplings to the environment leading to DFS, namely when the Lindblad operators do only depend on ghost operators which span the Hilbert space but commute with the symmetry $\opJ$, as explicitly demonstrated with the Heisenberg XXZ spin model.
In terms of the eigenoperator, DFS exist if $[\opL_{\alpha} , \opA_{nm}]  \,{=}\, [\opL_{\alpha}^{\dagger},\opA_{nm}]  \,{=}\, 0$~\cite{Buca2019}. Note, however, that the latter condition is not necessary since for the adjoint Lindbladian to have real eigenvalues, it follows from~\eqref{eq:adj-parts-L} that only the sum
$\opL_{\alpha}^{\dagger} [\opA,\opL_{\alpha} ] {+} [ \opL^{\dagger}_{\alpha}, \opA]\opL_{\alpha}  \,{=}\, 0$ must vanish. How this extends the classes of possible interactions with the environment leading to DFS even further beyond those depending only on ghost operators will be a subject of further research \cite{in_preparation}.

\paragraph*{Acknowledgments:}
JD acknowledges Andreas Buchleitner, Charlie Duclut, Christian Johansen and Fran\-cesco Piazza for helpful discussions.


\def\articletitle#1{\emph{#1.}}

\onecolumngrid
\newpage

\def\textfraction{0.0}
\def\topfraction{1.0}
\def\bottomfraction{1.0}

\makeatletter 
\renewcommand{\thefigure}{S\@arabic\c@figure}
\renewcommand{\theequation}{S\@arabic\c@equation}
\makeatother

\fboxsep=1pt

\definecolor{gray}{rgb}{0.5,0.5,0.5}
\definecolor{green}{rgb}{0,0.4,0.2}

\definecolor{darkblue}{rgb}{0,0,0.5}
\definecolor{darkred}{rgb}{0.5,0,0}
\definecolor{darkgreen}{rgb}{0,0.5,0}

\def\us#1{\color{darkblue}[us] #1 \color{black}}
\def\jm#1{\color{darkred}[jm] #1 \color{black}}
\def\jd#1{\color{darkgreen}[jd] #1 \color{black}}
\def\out#1{\color{gray}#1\color{black}}

\def\d{{\rm d}}
\def\e#1{{\rm e}^{#1}}
\def\i{{\rm i}}
\def\tr{{\rm tr}}
\def\del#1{\delta\hspace{-0.05em}#1}

\def\cL{\mathcal{L}}

\def\Sup{{\color{red} \uparrow}}
\def\Sdo{{\color{blue} \downarrow}}

\def\dpot{D}
\def\lpot{L}
\def\sd{s}

\def\op#1{\mathbb{#1}}
\def\opH{\op{H}}
\def\opL{\op{L}}
\def\opA{\op{A}}
\def\opJ{\op{J}}
\def\opS{\op{S}}
\def\opF{\op{F}}
\def\opG{\op{G}}
\def\opD{\op{D}}



\def\refname{{\normalsize References}}
\def\refname{\vspace*{-10mm}}

\def\Eb{E_\mathrm{b}}

\def\headrulewidth{0pt}
\thispagestyle{plain}

\begin{center}\bfseries\large
Symmetry-induced decoherence-free subspaces \\ --\,Supplemental material\,--\\[2mm]
\rm\large
Jonathan Dubois, Ulf Saalmann and Jan Michael Rost
\end{center}

\bigskip
\section{Lindbladian operator, its adjoint operator and its semiclassical limit}

In the quantum mechanical context, the inner product between two observables $\opF$ and $\opG$ in the Hilbert space is given by $\langle \opF,\opG\rangle = \tr \; ( \opF^{\dagger} \opG )$ where $\opF^{\dagger}$ denotes the complex-conjugate transpose of $\opF$. The adjoint  $\mathcal{L}^{\dagger}$ of the Lindbladian is defined through $\langle\cL^{\dagger} \opF , \opG \rangle = \langle \opF ,\cL \opG \rangle$. The Lindbladian and its adjoint are given by~\cite{Albert2014}
\begin{subequations}\begin{align}
   \cL \rho & = \dfrac{1}{\hbar} [ \opH , \rho] + \dfrac{\i}{\hbar} \sum_{\alpha} \left( [ \opL_{\alpha} , \rho \opL^{\dagger}_{\alpha}  ] + [ \opL_{\alpha} \rho , \opL_{\alpha}^{\dagger}  ]   \right) , \\
   \cL^{\dagger} \opF & = \dfrac{1}{\hbar} [\opH,\opF] - \dfrac{\i}{\hbar} \sum_{\alpha} \left( \opL_{\alpha}^{\dagger} [\opF,\opL_{\alpha}] + [\opL^{\dagger}_{\alpha},\opF] \opL_{\alpha} \right) ,
\end{align}\end{subequations}
respectively, such that we have $\dot{\rho} = - \i\cL \rho$ and $\dot{\opF} = \i\cL^{\dagger} \opF$.
In the limit $\opL_{\alpha}{=}0$ it is $\mathcal{L}{=}\mathcal{L}^{\dagger}$. 
In the semiclassical limit, the scalar product in the Hilbert space with phase-space variables $\mathbf{z}$ is given by $\langle F,G\rangle = \int\!\d\rule{-0.05em}{1.9ex}^{n}\!\!z\, F^{*}(\mathbf{z}) G(\mathbf{z})$ for an $n$-dimensional phase space. 
The semiclassical limit of the Lindbladian and its adjoint are given by~\cite{Dubois2021}
\begin{subequations}\begin{align}
   \cL \rho & = \i \lbrace H , \rho \rbrace - \sum_{\alpha} \Big( \left( \lbrace L_{\alpha} , \rho L_{\alpha}^{*} \rbrace + \lbrace L_{\alpha} \rho , L_{\alpha}^{*}  \rbrace  \right) - \dfrac{\i \hbar}{2} \left( \lbrace \lbrace L_{\alpha},\rho \rbrace , L_{\alpha}^{*} \rbrace + \lbrace L_{\alpha} , \lbrace \rho , L_{\alpha}^{*} \rbrace \rbrace \right) \Big) , \\
   \cL^{*} F & = \i \lbrace H,F\rbrace + \sum_{\alpha} \Big( \left( L_{\alpha}^{*} \lbrace F,L_{\alpha} \rbrace + \lbrace L_{\alpha}^{*} , F \rbrace L_{\alpha} \right) + \dfrac{\i \hbar}{2} \left( \lbrace  \lbrace L_{\alpha},F \rbrace , L_{\alpha}^{*} \rbrace  + \lbrace L_{\alpha} , \lbrace F,L_{\alpha}^{*} \rbrace \rbrace \right) \Big) . \label{eq:Lindbladian_adjoint_semiclassical}
\end{align}\end{subequations}
The Lindbladian can always be cast into a Fokker-Planck equation~\cite{Dubois2021,Strunz1998}. 

\section{Heisenberg spin model}

We provide details of the classical construction of ghost variables in order to define environments that allow for decoherence-free subspaces for the Heisenberg  model presented in the text.

\subsection*{Classical spin algebra}

We consider a chain of $N$ spins with in the classical limit is described by the spin variables $\mathbf{S}_i = S^x_i \mathbf{e}_x + S^y_i \mathbf{e}_y + S^z_i \mathbf{e}_z$ with $i=1,...,N$ and the spin algebra
\begin{equation}
  \label{eq:PB_spinalgebra}
    \lbrace S^{\alpha}_i , S^{\beta}_j \rbrace = \delta_{ij} \varepsilon_{\alpha \beta \gamma} S^{\gamma}_i 
\end{equation}
with $\varepsilon_{\alpha \beta \gamma}$ the Levi-Civita symbol. The corresponding non-canonical Poisson bracket is given by
\begin{subequations}
\label{eq:Poisson_bracket_spin}
\begin{equation}
    \lbrace F , G \rbrace = \sum_{i=1}^{N} \mathbf{S}_i \cdot \dfrac{\partial F}{\partial \mathbf{S}_i} \times \dfrac{\partial G}{\partial \mathbf{S}_i} .
\end{equation}
Equivalently, we can describe the dynamics with the beating variables $S^{\pm}_{i} = S^x_i \pm \i S^y_i$, for which the Poisson bracket reads
\begin{align}
    \lbrace F , G \rbrace = {} & \sum_{i=1}^{N} \bigg( {-} 2\i S^z_i \bigg( \dfrac{\partial F}{\partial S^+_i} \dfrac{\partial G}{\partial S^-_i} - \dfrac{\partial F}{\partial S^-_i} \dfrac{\partial G}{\partial S^+_i}  \bigg) 
    \notag\\ & \qquad
    + \i\, S^+_i \bigg( \dfrac{\partial F}{\partial S^+_i} \dfrac{\partial G}{\partial S^z_i} - \dfrac{\partial F}{\partial S^z_i} \dfrac{\partial G}{\partial S^+_i}  \bigg) - \i\, S^-_i \bigg( \dfrac{\partial F}{\partial S^-_i} \dfrac{\partial G}{\partial S^z_i} - \dfrac{\partial F}{\partial S^z_i} \dfrac{\partial G}{\partial S^-_i}  \bigg) \bigg) ,
\end{align}
\end{subequations}
with the fundamental spin algebra $\lbrace S^z_i , S^{\pm}_i \rbrace = \mp \i S^{\pm}_i$ and $\lbrace S^+_i , S^-_i \rbrace = - 2 \i S^z_i$. 
Note that in both cases, the classical Poisson bracket is non-canonical and is related to the quantum Lie bracket by $[\opF,\opG] \equiv \i \hbar \lbrace F , G \rbrace$.

\bigskip
\subsection*{Cylindrical coordinates and conserved quantity}

In cylindrical coordinates $\{S_i^z,\theta_i\}$ it is $\mathbf{S}_i = \sqrt{S_i{}^2 {-} (S^z_i)^2}\, (\mathbf{e}_x \cos \theta_i + \mathbf{e}_y \sin \theta_i) + S^z_i \mathbf{e}_z$ or equivalently\\ $S^{\pm}_i = \sqrt{S_i{}^2 {-} (S_i^z)^2} \, \exp (\pm\i \theta)$ with $S_i{}^2=|\mathbf{S}_i|^2$ a Casimir invariant. 
Therewith the Hamiltonian (in the semiclassical limit) reads
\begin{equation}
    \label{eq:Heisenber_model_canonical_variables}
    H (\theta_i , S^z_i) = \sum_{i=1}^{N} \bigg( \omega_i S^z_i + \sum_{j=1}^{N}  \left( 2 \sigma_{ij} \sqrt{[S_i^2-(S^z_i)^2] [S_j^2-(S^z_j)^2]} \cos \left( \theta_i - \theta_j \right) + \Delta_{ij} S^z_i S^z_j \right)  \bigg) ,
\end{equation}
and the Poisson bracket becomes 
\begin{equation}
    \label{eq:Poisson_bracket_cylindrical_coordinates}
    \lbrace F , G \rbrace = \sum_{i=1}^{N} \bigg( \dfrac{\partial F}{\partial\!\theta_i} \dfrac{\partial G}{\partial S^z_i} - \dfrac{\partial F}{\partial S^z_i} \dfrac{\partial G}{\partial\!\theta_i} \bigg) .
\end{equation}
From the Hamiltonian \eqref{eq:Heisenber_model_canonical_variables}, it is clear that 
\begin{equation}
    \label{eq:J_Heisenberg_spin_model}
    J = \dfrac{1}{N} \sum_{i=1}^{N} S^z_i,
\end{equation}
is a conserved quantity due to the invariance under a rotation around $\mathbf{e}_z$.
The variable canonically conjugate to $J$ such that $\lbrace \theta , J \rbrace = 1$ is given by
\begin{equation}
    \theta = \sum_{i=1}^N \theta_i ,
\end{equation}
corresponding to the total angle. 
There is some freedom for choosing the set of canonically-conjugate variables from there and therefore the \emph{ghost variables}. Indeed, note that one can 
transform one set of ghost variables into another one by canonical transformations. The most natural starting point are the Jacobi coordinates for the celestial many-body problem  $\Delta \theta_i = \theta_{i+1}- \theta_{i}$. It is clear that $\lbrace \Delta \theta_i , J \rbrace = \lbrace \Delta \theta_i , \theta \rbrace = 0$ for $i=1,...,N-1$. The goal is now to find the set of variables $\widetilde{S}^z_i$ canonically conjugate to $\Delta \theta_i$, i.e.,  $\lbrace \Delta \theta_i , \widetilde{S}^z_j \rbrace = \delta_{ij}$ with $\delta_{ij}$ the Kronecker delta. 

\subsection*{Ghost variables using canonical transformations}
For doing so, we use the $F_2$ generating function~\cite{Goldstein1980}. We recall that for a set of canonically conjugate variables $(\boldsymbol{\theta},\mathbf{S}^z)$ the new set of canonically conjugate variables $(\widetilde{\boldsymbol{\theta}},\widetilde{\mathbf{S}}^z)$ can be found by using transformations, such that $F_2 (\boldsymbol{\theta},\widetilde{\mathbf{S}}^z)$, such that
\begin{equation}
    \label{eq:F2_canonical_transformation}
    \mathbf{S}^z = \dfrac{\partial F_2}{\partial \boldsymbol{\theta}} , \qquad \widetilde{\boldsymbol{\theta}} = \dfrac{\partial F_2}{\partial \widetilde{\mathbf{S}}^z}.
\end{equation}
Here, the old variables are given by $\boldsymbol{\theta} = (\theta_1, ... , \theta_{N})$ and $\mathbf{S}^z = (S^z_1, ... , S^z_{N})$. The new variables are $\widetilde{\boldsymbol{\theta}} = ( \Delta \theta_1 , ... , \Delta \theta_{N{-}1},\theta )$ and $\widetilde{\mathbf{S}}^z = ( \widetilde{S}^z_1 , ... , \widetilde{S}^z_{N{-}1},J)$. Given the form of the new variables $\widetilde{\boldsymbol{\theta}}$ which we have imposed, the generating function is given by
\begin{equation}
    \label{eq:F2_generating_function}
    F_2 (\boldsymbol{\theta} , \widetilde{\mathbf{S}}^z) = J \sum_{i=1}^{N} \theta_i + \sum_{i=1}^{N{-}1} \widetilde{S}^z_i (\theta_{i+1} - \theta_{i}).
\end{equation}
Indeed, from~\eqref{eq:F2_canonical_transformation} and~\eqref{eq:F2_generating_function}, we obtain that $\widetilde{\theta}_0 \equiv \theta =  \sum_{i=1}^{N} \theta_i$ and $\widetilde{\theta}_i \equiv \Delta \theta_i = \theta_{i+1} - \theta_{i}$ for $i = 1,...,N{-}1$. The expression of the old momenta with respect to the new ones are found using~\eqref{eq:F2_canonical_transformation} as
\begin{equation}
    \label{Sz_to_Sztilde}
    S^z_1 = J - \widetilde{S}^z_1 , \qquad S^z_i = J + \widetilde{S}^z_{i{-}1} - \widetilde{S}^z_{i} , \qquad S^z_{N} = J + \widetilde{S}^z_{N{-}1} , 
\end{equation}
for $i=2,...,N{-}1$. We can  easily check that~\eqref{eq:J_Heisenberg_spin_model} is fulfilled. In order to invert this transformation and obtain the expression of the ghost variables with respect to $S^z_i$, a convenient way is to write it in a matrix form
\def\onebymplusone{\dfrac{1}{M{+}1}}%
\def\onebymplusone{1/(M{+}1)}%
\begin{equation}
    \begin{bmatrix}
        S^z_1 \\ 
        S^z_2 \\ 
        S^z_3 \\ 
        \vdots \\
        S^z_{N{-}2} \\ 
        S^z_{N{-}1} \\ 
        S^z_{N} \\ 
    \end{bmatrix} =
    \begin{bmatrix}
        -1 & 0 & 0 & \hdots & 0 & 0 & 1 \\ 
        1 & -1 & 0 & \hdots & 0 & 0 & 1 \\ 
        0 & 1 & -1 &  \hdots & 0 & 0 & 1 \\ 
        \vdots & \vdots & \vdots & \hdots & \vdots & \vdots & \vdots \\
        0 & 0 & 0 & \hdots & -1 & 0 & 1 \\ 
        0 & 0 & 0 & \hdots & 1 & -1 & 1 \\ 
        0 & 0 & 0 & \hdots & 0 & 1 & 1\\ 
    \end{bmatrix}
    \begin{bmatrix}
        \widetilde{S}^z_1 \\ 
        \widetilde{S}^z_2 \\ 
        \vdots \\
        \widetilde{S}^z_{N{-}3} \\ 
        \widetilde{S}^z_{N{-}2} \\ 
        \widetilde{S}^z_{N{-}1} \\ 
        J \\ 
    \end{bmatrix} .
\end{equation}
The inverse transformation is given by
\begin{equation}
     \begin{bmatrix}
        \widetilde{S}^z_1 \\ 
        \widetilde{S}^z_2 \\ 
        \vdots \\
        \widetilde{S}^z_{N{-}3} \\ 
        \widetilde{S}^z_{N{-}2} \\ 
        \widetilde{S}^z_{N{-}1} \\ 
        J \\ 
    \end{bmatrix} =
    \dfrac{1}{N}
    \begin{bmatrix}
        -N{+}1 & 1 & 1 & \hdots & 1 & 1 & 1 \\ 
        -N{+}2 & -N{+}2 & 2 & \hdots  & 2 & 2 & 2 \\ 
        -N{+}3 & -N{+}3 & -N{+}3 & \hdots  & 3 & 3 & 3 \\ 
        \vdots & \vdots & \vdots & \hdots & \vdots & \vdots & \vdots \\
        -2 & -2 & -2 & \hdots & - 2 & N{-}2 & N{-}2 \\ 
        -1 & -1 & -1 & \hdots & -1 & -1 & N{-}1\\ 
        1 & 1 & 1 & \hdots & 1 & 1 & 1 \\ 
    \end{bmatrix}
    \begin{bmatrix}
        S^z_1 \\ 
        S^z_2 \\ 
        S^z_3 \\ 
        \vdots \\
        S^z_{N{-}2} \\ 
        S^z_{N{-}1} \\ 
        S^z_{N} \\ 
    \end{bmatrix} .
\end{equation}
Therefore, the ghost variables are given by the Jacobi momentum coordinates
\begin{equation}
    \widetilde{S}^z_k = \dfrac{k}{N} \sum_{i=1}^{N} S^z_i -\sum_{i=1}^{k} S^z_i, \qquad \Delta \theta_k = \theta_{k+1} - \theta_{k} ,
    \qquad k = 1, ...,N{-}1
\end{equation}
as given in the main text. One can easily check that the transformation is canonical with $\lbrace \theta , J \rbrace = 1$, $\lbrace \Delta \theta_i , \widetilde{S}^z_j \rbrace = \delta_{ij}$ and $\lbrace \theta , \widetilde{S}^z_i \rbrace = \lbrace J , \widetilde{S}^z_i \rbrace = \lbrace \theta , \Delta \theta_i \rbrace = \lbrace J , \Delta \theta_i \rbrace = \lbrace \widetilde{S}^z_i , \widetilde{S}^z_j \rbrace = \lbrace \Delta \theta_i , \Delta \theta_j \rbrace = 0$ for all $i$ and $j$. 
In the same way, we find the change of coordinates for the angles and relative angles. 

In this new set of variables, the Hamiltonian $H(J , \Delta \theta_i , \widetilde{S}_i^z)$ replacing the one in Eq.~\eqref{eq:Heisenber_model_canonical_variables} is independent of the angle $\theta$. 

\subsection*{Beating variables and transformation back to Cartesian spin algebra}

As mentioned in the main text, from canonical variables associated to a conserved quantity $(\theta,J)$, one can construct beating observables $A = f(J) \exp ( {-} \i \theta )$ and $A^{\dagger} = f(J) \exp ( \i \theta )$. 
One can always perform the inverse change of coordinates from~\eqref{eq:Poisson_bracket_cylindrical_coordinates} to~\eqref{eq:Poisson_bracket_spin} by going back to a spin algebra. For instance,  using $f(J)=1$  we find the beating variables
\begin{equation}
    A = \prod_{i=1}^N \dfrac{S^{-}_i}{\sqrt{S_i{}^2 - (S_i^z)^2}} , \qquad A^{*} = \prod_{i=1}^N \dfrac{S^{+}_i}{\sqrt{S_i{}^2 - (S_i^z)^2}} .
\end{equation}
In the previous paragraph, we have described the canonical transformation from $(\theta_i , S_i^z)$ to variables $(\theta , J , \Delta \theta_i , \widetilde{S}_i^z)$. One can always go back to a spin representation by performing the inverse change of coordinates from~\eqref{eq:Poisson_bracket_cylindrical_coordinates} to~\eqref{eq:Poisson_bracket_spin}. Here, this change of coordinates reads 
\begin{subequations}\label{eq:can2noncanPB}\begin{align}
    \widetilde{S}^{\pm}_N & = \sqrt{\widetilde{S}_N{}^2 - J^2} \exp (\pm \i \theta) , & \widetilde{S}^z_N & = J , \\
    \widetilde{S}^{\pm}_i & = \sqrt{\widetilde{S}_i{}^2 - (\widetilde{S}^z_i)^2} \exp (\pm \i \Delta\theta_i) ,  & \widetilde{S}^z_i & = \widetilde{S}_i^z ,
     & i & = 1, ...,N{-}1,
\end{align}\end{subequations}
which allows us to obtain the spin algebra~\eqref{eq:Poisson_bracket_cylindrical_coordinates}. 
Note that $A \propto \widetilde{S}_N^{-}$ and $A^{*}  \propto \widetilde{S}_N^{+}$ corresponding to the beating variables associated with the conserved quantity $J$ introduced in the main text.

To summarize, we  started with a Hamiltonian in terms of  a spin algebra in the main text. Switching to canonical variables we obtained Hamiltonian~\eqref{eq:Heisenber_model_canonical_variables}. Then, using a canonical transformation, we found the new coordinates which consist of the DOFs of the collective spin variables $(\theta, J)$ and the ghost variables $(\Delta \theta_i , \widetilde{S}_i^z)$.
From~\eqref{eq:can2noncanPB}, the expression of the ghost variables with respect to the initial spin variables read by means of the abbreviations
$S_i^\rho\equiv\sqrt{S_i^2 - (S_i^z)^2}$ and $\widetilde{S}_i^\rho\equiv\sqrt{\widetilde{S}_i^2 - (\widetilde{S}_i^z)^2}$ (with Casimir invariants $\widetilde{S}_i^2 = |\widetilde{\mathbf{S}}_i|^2$)
\begin{subequations}\label{eq:Heisenberg_classical} \begin{align}
  \widetilde{S}^z_N & = \dfrac{1}{N} \sum_{i=1}^{N} S_i^z,
  &
    \widetilde{S}^{\pm}_N & = \frac{\widetilde{S}_N^\rho}{\prod_{i=1}^{N} S_i^\rho} \prod_{i=1}^{N} S_i^{\pm}, \label{eq:Heisenberg_A_classical} 
    \\
  \widetilde{S}^z_{k} &= \dfrac{k}{N} \sum_{i=1}^{N} S^z_i -\sum_{i=1}^{k}S_i^z,
     &
    \widetilde{S}^{\pm}_{k} & =  \dfrac{\widetilde{S}_k^\rho}{S_{k{+}1}^\rho\,S_{k}^\rho} S^{\pm}_{k{+}1} S^{\mp}_{k},
    & k & = 1, ...,N{-}1.
    \label{eq:Heisenberg_GO_classical}
\end{align}\end{subequations}
This transformation is canonical, in the sense that it preserves the form of the Poisson bracket~\eqref{eq:PB_spinalgebra} for the spin algebra. 
The beating variables associated with the conserved quantity $J$ are therefore $A = \widetilde{S}_{N}^{-}$. We go from $\lbrace S^i_n , S^j_m \rbrace = \varepsilon_{ijk} S_m^k \delta_{nm}$, to $\lbrace \widetilde{S}^i_n , \widetilde{S}^j_m \rbrace = \varepsilon_{ijk} \widetilde{S}_m^k \delta_{nm}$. Or equivalently, from $\lbrace S^{\pm}_n , S^z_m \rbrace = \pm \i S^{\pm}_n \delta_{nm}$ and $\lbrace S^{+}_n , S^{-}_m \rbrace = - 2 \i S^z_n \delta_{nm}$ to $\lbrace \widetilde{S}^{\pm}_n , \widetilde{S}^z_m \rbrace = \pm \i \widetilde{S}^{\pm}_n \delta_{nm}$ and $\lbrace \widetilde{S}^{+}_n , \widetilde{S}^{-}_m \rbrace = - 2 \i \widetilde{S}^z_n \delta_{nm}$.
Therefore, the new spin variables $\widetilde{\mathbf{S}}_i$ for $i=1,...,N{-}1$ are \emph{ghost variables} which do not affect the DOF associated with the conserved quantity $\widetilde{\mathbf{S}}_N$.

\subsection*{Ghost and beating operators}

From~\eqref{eq:Heisenberg_A_classical}, we upgrade the beating and ghost variables to their quantum mechanical counterparts, the beating and ghost operators. To do so, we replace the variables by their associated operators (which are matrices) in the expressions~\eqref{eq:Heisenberg_A_classical}. 
We first notice that the quantities $\opS_i^{\rho}$, in the basis of the operators $\opS_i^z$, are diagonal matrices since $\opS_i$ is proportional to identity (Casimir invariant) and $\opS_i^z$ is diagonal. This allows us to compute the square roots and inverse of the operators $\opS_i^{\rho}$. In addition, these operators commute with each other. 
For the Heisenberg spin model, we find a simplified version of a beating operator 
\begin{equation}
  \opA = \prod_{i{=}1}^{N} \opS_i^{-},
\end{equation}
since these operators are defined up to a factor.
The expectation value of this operator oscillates in time.
From the classical ghost variables~\eqref{eq:Heisenberg_GO_classical}, we obtain the ghost operators in the main text 
\begin{align}
  \widetilde{\opS}^z_{k} &= \dfrac{k}{N} \sum_{i{=}1}^{N} \opS^z_i -\sum_{i{=}1}^{k} \opS_j^z
     &
    \widetilde{\opS}^{\pm}_{k} & =  \opD_k \; \opS^{\pm}_{k{+}1} \opS^{\mp}_{k}, \qquad \opD_k = \dfrac{\widetilde{\opS}_k^\rho}{\opS_{k{+}1}^\rho\,\opS_{k}^\rho} . 
\end{align}
As mentioned above, all $\opS_k^\rho$ are diagonal, which makes the expression for $\opD_k$ unique.

\subsection*{Parameters for the Heisenberg model presented in the text}

In Figs.\,1 and~2, the self-energy $\omega_i$ of each spin site $i$ has been generated randomly. 
The actual values and the matrices for nearest-neighbors interactions, cf.\ Eq.\,(6) in the text, are
\begin{equation}
    [\omega_i] = 
    \begin{bmatrix}
    0.6358 \\ 0.9452 \\ 0.2089 \\ 0.7093 \\ 0.2362 \\ 0.1194 \\
    \end{bmatrix} , \quad 
    [\sigma_{ij}] = \frac{1}{2}
    \begin{bmatrix}
    0 & 1 & 0 & 0 & 0 & 1 \\
    1 & 0 & 1 & 0 & 0 & 0 \\
    0 & 1 & 0 & 1 & 0 & 0 \\
    0 & 0 & 1 & 0 & 1 & 0 \\
    0 & 0 & 0 & 1 & 0 & 1 \\
    1 & 0 & 0 & 0 & 1 & 0 \\
    \end{bmatrix} , 
    \quad 
    [\Delta_{ij}] = \frac{1}{4}
    \begin{bmatrix}
    0 & 1 & 0 & 0 & 0 & 1 \\
    1 & 0 & 1 & 0 & 0 & 0 \\
    0 & 1 & 0 & 1 & 0 & 0 \\
    0 & 0 & 1 & 0 & 1 & 0 \\
    0 & 0 & 0 & 1 & 0 & 1 \\
    1 & 0 & 0 & 0 & 1 & 0 \\
    \end{bmatrix} .
\end{equation}
For the Lindblad operators in Fig.\,2c, we have used a combination of ghost operators such that $\opL_{\alpha_{ij}} = \gamma (\opS_i^z - \opS_j^z)$. This is done using~\eqref{Sz_to_Sztilde}.

\bigskip\noindent
As initial condition we have taken
$\rho (0) = \ket{\psi} \bra{\psi}$ with 
\begin{equation}\label{eq:inital_state}
\ket{\psi} = \frac{\sum_{k_1\ldots k_6}\ket{k_1\ldots k_6}C_{k_1\ldots k_6}}
{\sqrt{\sum_{k_1\ldots k_6}C_{k_1\ldots k_6}{}^2}}
\quad\mbox{where}\quad C_{k_1\ldots k_6} =
\!\sum_{\xi=\pm63}\!\exp\big({-}\big[\mbox{$\sum_j$} k_j\,2^{j}{-}\xi\big]^2/128\big)
\end{equation}
and $k_j=\{{-}\frac{1}{2},{+}\frac{1}{2}\}$ and thus 64 different states $\ket{k_1\ldots k_6}$.


\begin{thebibliography}{10}
\bibitem{Eins}
The unit operators in Eqs.\,(\ref{eq:Lindbladian_block}a,b) follow the
  definition for $\opH_n$ and $\opL_{\alpha n}$. Thus $\opone_n \,{=}\,{\rm
  tr}_n\opP_n \opone \opP_n \,{=}\,{\rm tr}_n\opP_n$ is a unit operator of
  dimension $\eta_n$.
\end{thebibliography}

\vfill
\def\articletitle#1{\emph{#1.}}

\end{document}